\documentclass[aps,prc,twocolumn,superscriptaddress,showpacs,showkeys]{revtex4}
\usepackage{hyperref}

\usepackage{graphicx}
\usepackage{dcolumn}
\usepackage{bm}

\usepackage{float}

\usepackage[usenames]{color}

\begin{document}


\title{Comparison of nonequilibrium processes in p+Ni and p+Au collisions at GeV energies}



\author{A.Budzanowski}
\affiliation{H. Niewodnicza{\'n}ski Institute of Nuclear Physics
PAN, Radzikowskiego 152, 31342 Krak{\'o}w, Poland}
\author{M.Fidelus}
\affiliation{M. Smoluchowski Institute of Physics, Jagellonian
University, Reymonta 4, 30059 Krak{\'o}w, Poland}
\author{D.Filges}
\affiliation{Institut f{\"u}r Kernphysik, Forschungszentrum
J{\"u}lich, D-52425 J{\"u}lich, Germany}
\author{F.Goldenbaum}
\affiliation{Institut f{\"u}r Kernphysik, Forschungszentrum
J{\"u}lich, D-52425 J{\"u}lich, Germany}
\author{H.Hodde}
\affiliation{Institut f{\"ur} Strahlen- und Kernphysik, Bonn
University,  D-53121 Bonn, Germany}
\author{L.Jarczyk}
\affiliation{M. Smoluchowski Institute of Physics, Jagellonian
University, Reymonta 4, 30059 Krak{\'o}w, Poland}
\author{B.Kamys}  \email[Corresponding author: ]{ufkamys@cyf-kr.edu.pl}
\affiliation{M. Smoluchowski Institute of Physics, Jagellonian
University, Reymonta 4, 30059 Krak{\'o}w, Poland}
\author{M.Kistryn}
\affiliation{H. Niewodnicza{\'n}ski Institute of Nuclear Physics
PAN, Radzikowskiego 152, 31342 Krak{\'o}w, Poland}
\author{St.Kistryn}
\affiliation{M. Smoluchowski Institute of Physics, Jagellonian
University, Reymonta 4, 30059 Krak{\'o}w, Poland}
\author{St.Kliczewski}
\affiliation{H. Niewodnicza{\'n}ski Institute of Nuclear Physics
PAN, Radzikowskiego 152, 31342 Krak{\'o}w, Poland}
\author{A.Kowalczyk}
\affiliation{M. Smoluchowski Institute of Physics, Jagellonian
University, Reymonta 4, 30059 Krak{\'o}w, Poland}
\author{E.Kozik}
\affiliation{H. Niewodnicza{\'n}ski Institute of Nuclear Physics
PAN, Radzikowskiego 152, 31342 Krak{\'o}w, Poland}
\author{P.Kulessa}
\affiliation{H. Niewodnicza{\'n}ski Institute of Nuclear Physics
PAN, Radzikowskiego 152, 31342 Krak{\'o}w, Poland}
\affiliation{Institut f{\"u}r Kernphysik, Forschungszentrum
J{\"u}lich, D-52425 J{\"u}lich, Germany}
\author{H.Machner}
\affiliation{Institut f{\"u}r Kernphysik, Forschungszentrum
J{\"u}lich, D-52425 J{\"u}lich, Germany}
\author{A.Magiera}
\affiliation{M. Smoluchowski Institute of Physics, Jagellonian
University, Reymonta 4, 30059 Krak{\'o}w, Poland}
\author{B.Piskor-Ignatowicz}
\affiliation{M. Smoluchowski Institute of Physics, Jagellonian
University, Reymonta 4, 30059 Krak{\'o}w, Poland}
\affiliation{Institut f{\"u}r Kernphysik, Forschungszentrum
J{\"u}lich, D-52425 J{\"u}lich, Germany}
\author{K.Pysz}
\affiliation{H. Niewodnicza{\'n}ski Institute of Nuclear Physics
PAN, Radzikowskiego 152, 31342 Krak{\'o}w, Poland}
\affiliation{Institut f{\"u}r Kernphysik, Forschungszentrum
J{\"u}lich, D-52425 J{\"u}lich, Germany}
\author{Z.Rudy}
\affiliation{M. Smoluchowski Institute of Physics, Jagellonian
University, Reymonta 4, 30059 Krak{\'o}w, Poland}
\author{R.Siudak}
\affiliation{H. Niewodnicza{\'n}ski Institute of Nuclear Physics
PAN, Radzikowskiego 152, 31342 Krak{\'o}w, Poland}
\affiliation{Institut f{\"u}r Kernphysik, Forschungszentrum
J{\"u}lich, D-52425 J{\"u}lich, Germany}
\author{M.Wojciechowski}
\affiliation{M. Smoluchowski Institute of Physics, Jagellonian
University, Reymonta 4, 30059 Krak{\'o}w, Poland}

\collaboration{PISA - \textbf{P}roton \textbf{I}nduced
\textbf{S}p\textbf{A}llation collaboration}

\date{\today}

\begin{abstract}
The energy and angular dependence of double differential cross
sections $d^{2}\sigma/d\Omega dE$ were measured for
$p,d,t,^{3,4,6}$He, $^{6,7,8}$Li, $^{7,9,10}$Be, $^{10,11}$B, and C
produced in collisions of  1.2, 1.9, and 2.5 GeV protons with a Ni
target. The shape of the spectra and angular distributions does
almost not change  whereas the  absolute value of the cross sections
increases by a factor $\sim$ 1.7 for all ejectiles in this beam
energy range. It was found that energy and angular dependencies of
the cross sections cannot be reproduced by the microscopic model of
intranuclear cascade with coalescence of nucleons and the
statistical model for evaporation of particles from excited,
equilibrated residual nuclei. The inclusion of nonequilibrium
processes, described by a phenomenological model of the emission
from fast and hot moving sources, resulting from break-up of the
target nucleus by impinging proton, leads to very good reproduction
of data. Cross sections of these processes are quite large,
exhausting approximately half of the total production cross
sections. Due to good reproduction of energy and angular
dependencies of $d^{2}\sigma/d\Omega dE$ by model calculation it was
possible to determine total production cross sections for all
studied ejectiles.  Results obtained in this work point to the
analogous reaction mechanism for proton induced reactions on Ni
target as that observed previously for  Au target in the same beam
energy range.

\end{abstract}



\maketitle


\section{\label{sec:introduction} Introduction}

The recent successful analysis \cite{BUB07A,BUD08A} of the inclusive
spectra and angular distributions of double differential cross
sections $d^{2}\sigma/d\Omega dE$ for light charged particles
(LCP's), i.e. particles with Z $\le$ 2, and intermediate mass
fragments (IMF's), i.e. ejectiles heavier than $^4$He, produced in
proton - Au collisions at proton beam energies 1.2, 1.9, and  2.5
GeV indicated, that the competition of two different processes is
essential for understanding of the mechanism of proton - gold
nucleus collisions. The traditional picture of the spallation
reactions, namely the intranuclear cascade of nucleon - nucleon
interactions followed by evaporation of particles from an
equilibrated residuum of the cascade is not able to reproduce
satisfactorily the experimental cross sections.  This fact leads to
the conclusion that a significant contribution of nonequilibrium
processes to the reaction mechanism is present.

It was found that the difference between the data and theoretical
cross sections from the above two-step model varies smoothly with
the scattering angle and with the energy of the ejectile.  It turned
out, that this variation may be well reproduced assuming the
isotropic emission  of nucleons and composite particles in the rest
frame of two or three sources moving forward, i.e., along the beam
direction. Such an effect has been interpreted in Refs.
\cite{BUB07A,BUD08A} as indication of break-up of the target nucleus
by fast proton from the beam.  In that physical picture the proton
drills a cylindrical hole in the nucleus knocking out the small
group of nucleons placed on the straight way of the proton through
the nucleus. This group moves quickly in forward direction behaving
as a fireball, which on its part emits nucleons and LCPs whereas the
"wounded" nucleus may decay into two excited prefragments which also
serve as sources of LCPs and IMFs.

The present study was performed with the aim to investigate, whether
the reaction mechanism observed and described in detail in Ref.
\cite{BUD08A} for Au target is also realized in collisions of
protons with other target nuclei. The Ni target was used for this
purpose because it has quite different properties than the Au
target.  The Ni nucleus is more than three times lighter, its N/Z
ratio ($\sim$ 1.1) is approximately 1.4 times smaller from that of
the Au nucleus (N/Z $\sim$ 1.5), and its binding energy ($\sim$ 8.8
MeV/nucleon) is significantly larger than that of the Au nucleus
($\sim$ 7.9 MeV/nucleon). The appearance of the same reaction
mechanism for such different target nuclei might suggest that this
phenomenon is of a general character, i.e., it occurs for all target
nuclei in the studied energy range.

To facilitate the comparison of the results from the present study
of the reactions in p+Ni system with results of previous
investigations concerning p+Au system \cite{BUD08A}, the present
paper is organized in similar way as reference \cite{BUD08A}.
Experimental data are discussed in the next section, the theoretical
analysis is described in the third section starting from IMF data
and followed by analysis of LCPs cross sections, discussion of
results are presented in the fourth and summary with conclusions in
the fifth section.


\section{\label{sec:experiment}Experimental data}

The experiment was performed with the selfsupporting Ni target of
the thickness  about of 150 $\mu$g/cm$^{2}$, irradiated by internal
proton beam of COSY (COoler SYnchrotron) of the J\"ulich Research
Center. The experimental setup and procedure of data taking were in
details described in Refs. \cite{BUB07A} and \cite{BAR04A}. The beam
was operated in so called supercycle mode to assure identical
experimental conditions for all three studied proton energies - 1.2,
1.9 and 2.5 GeV, i.e., the same setup, electronics, the target
thickness and its position.  In this mode several cycles were
alternated for each requested beam energy, consisting of protons
injection from JULIC cyclotron to COSY ring, their acceleration with
the beam circulating in the ring below the target, and irradiating
the target by slow movement of the beam in the upward direction. The
using of supercycle mode minimizes systematic effects which might
distort the studied energy dependence of the cross sections.

Double differential cross sections $d^{2}\sigma/d\Omega dE$ were
measured at seven scattering angles; 16$^0$, 20$^0$, 35$^0$, 50$^0$,
65$^0$, 80$^0$, and 100$^0$ as a function of energy of ejectiles for
the following isotopes $^{1,2,3}$H, $^{3,4,6}$He, $^{6,7,8}$Li,
$^{7,9,10}$Be, and $^{10,11}$B. The carbon ejectiles were only
charge identified.

The absolute normalization of the cross sections was achieved by
comparing the total production cross sections of $^7$Be particles,
obtained by angle and energy integration of measured differential
cross sections, with values of $^7$Be total production cross
sections published in the compilation of Bubak et al. \cite{BUB04A}.
It turned out that present values of total production cross sections
at proton beam energy 1.2 GeV are in perfect agreement with those
obtained by Herbach \emph{et al.} \cite{HER06A}.  The same quality
agreement was observed for total production cross sections of the
present work at 2.5 GeV and Li, Be, and B cross sections measured by
Raisbeck \emph{et al.} \cite{RAI75A} on Ni target with 3.0 GeV
proton beam.

\begin{figure}[ht]
\begin{center}
\includegraphics[angle=0,width=0.45\textwidth]{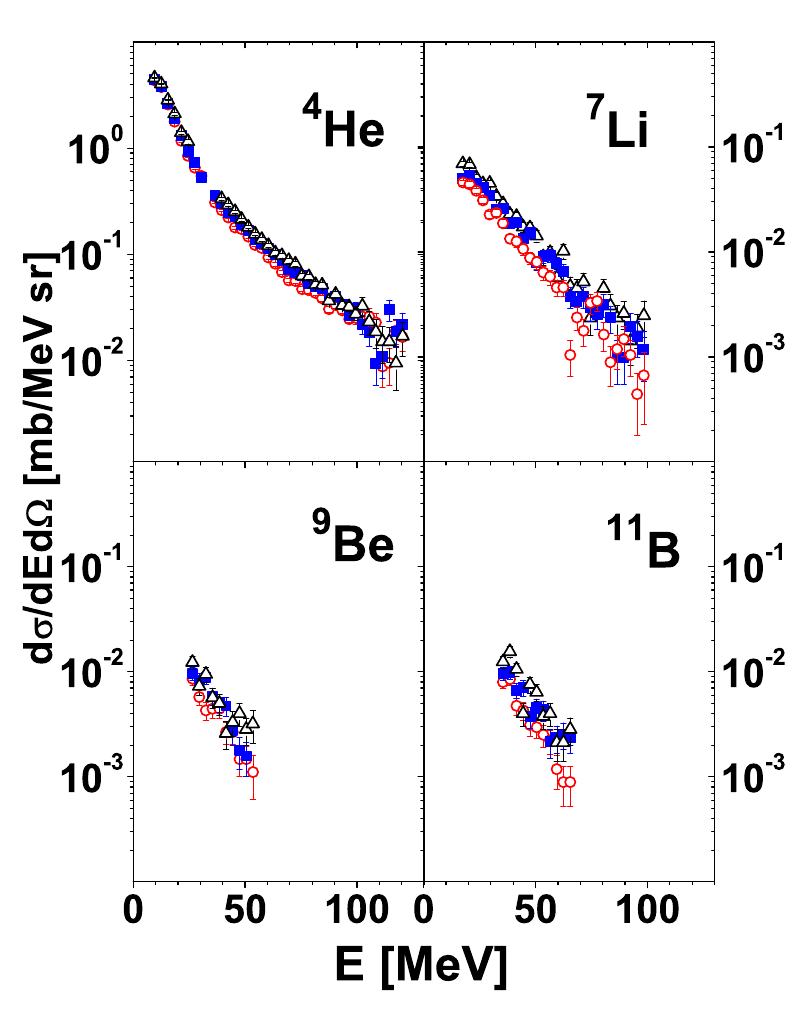}
\caption{\label{fig:helibeb}  Typical spectra of $^{4}$He, $^{7}$Li,
$^{9}$Be, and $^{11}$B ejectiles  (upper left, upper right, lower
left, and lower right parts of the figure, respectively) measured at
35$^{\circ}$ for three energies of the proton beam; 1.2, 1.9, and
2.5 GeV, impinging on to the Ni target. Open circles represent the
lowest energy, full squares -  the intermediate energy, and the open
triangles show the data for the highest energy. }
\end{center}
\end{figure}

 Typical spectra of isotopically
identified ejectiles obtained in the present experiment are shown in
Fig. \ref{fig:helibeb}. All spectra are smooth and do not change
their shapes with increasing beam energy, however, the magnitude of
the cross sections increases slightly with the energy.

\section{\label{sec:analysis} Theoretical analysis}

The analysis of the present experimental data was performed
according to the same procedure as that applied previously to the
data from proton induced reactions on the Au target in the work of
Budzanowski \emph{et al.} \cite{BUD08A}. First, the cross sections
were evaluated from the intranuclear cascade  with inclusion of
possibility to coalesce the outgoing nucleons into LCPs, and with
possibility to evaporate the particles from the excited,
equilibrated residua of intranuclear cascade. Such two-step model of
the reaction mechanism is most frequently used in the literature for
the description of spallation reactions at high proton energies. The
INCL4.3 computer program of Boudard \emph{et al.} \cite{BOU04A} has
been used for calculations of intranuclear cascade and GEM2 program
of S. Furihata \cite{FUR00A,FUR02A} has been applied to evaluate
evaporation cross sections. Since the data were generally
underestimated by two-step model a phenomenological analysis was
performed in the following.
This analysis was based on the
%
\begin{figure}[ht]
\begin{center}
\hspace*{-0.5cm}
\includegraphics[angle=0,width=0.5\textwidth]{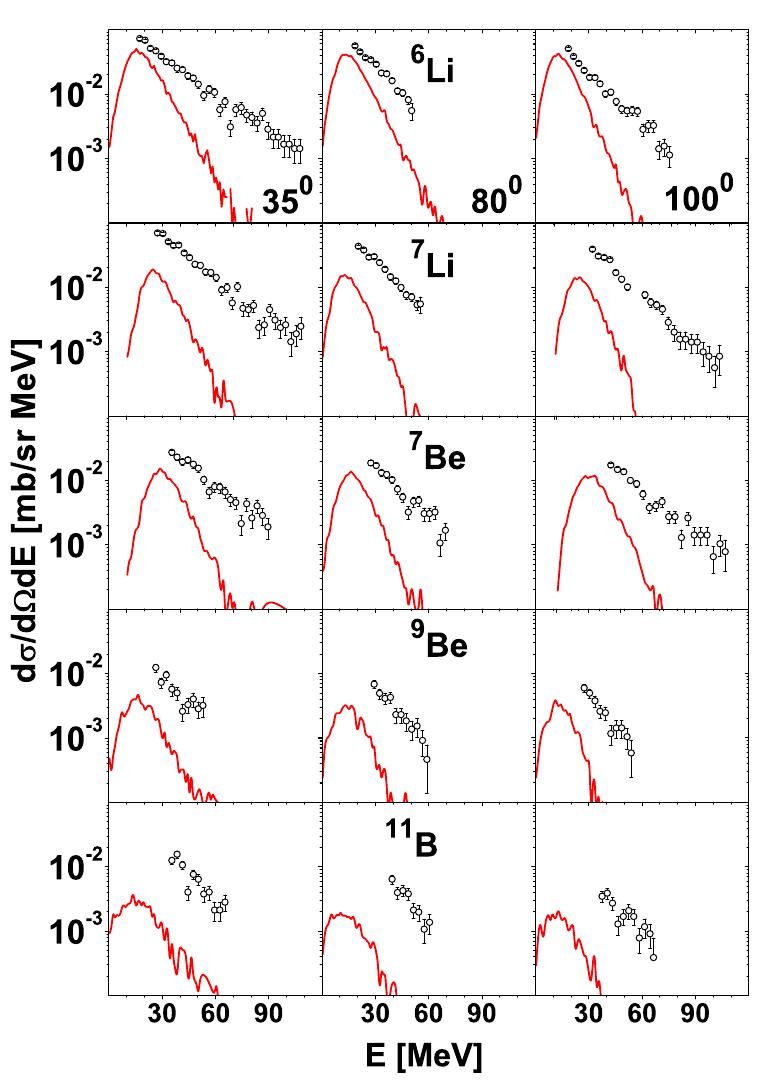}
\caption{\label{fig:libebinc}  Typical spectra of selected lithium,
beryllium, and boron isotopes from p+Ni collisions measured at
35$^{\circ}$, 80$^{\circ}$, and 100$^{\circ}$ (left, middle, and
right columns, respectively)  for 2.5 GeV proton beam impinging on
to the Ni target. The detected particles are listed in the central
panel of each row of pictures. Open circles represent the
experimental data,  and solid lines correspond to intranuclear
cascade followed by evaporation of particles, respectively. }
\end{center}
\end{figure}
%
assumption that additional processes exist besides the mechanisms
described above.  They were parameterized by incoherent sum of
isotropic emission of particles from highly excited sources moving
in forward direction, i.e., along to the beam. Each of the sources
has a Maxwellian distribution of the energy $E$ available for the
two body decay resulting in emission of the detected particles; $d^2
\sigma/dE d\Omega \sim \sqrt{E} \exp(-E/T)$. The velocity of
 the source - $\beta$ (in units of speed of light), its temperature
- $T$ (in MeV), and the contribution to the total production cross
section - $\sigma$ are treated as free parameters. The presence of
the Coulomb barrier, which hinders emission of low energy particles,
was taken into account multiplying the Maxwellian energy
distribution by a smooth function P(E) corresponding to the
transmission probability through the barrier.
%
\begin{figure}[ht]
\begin{center}
\vspace*{-0.25cm} \hspace*{-0.5cm}
\includegraphics[angle=0,width=0.5\textwidth]{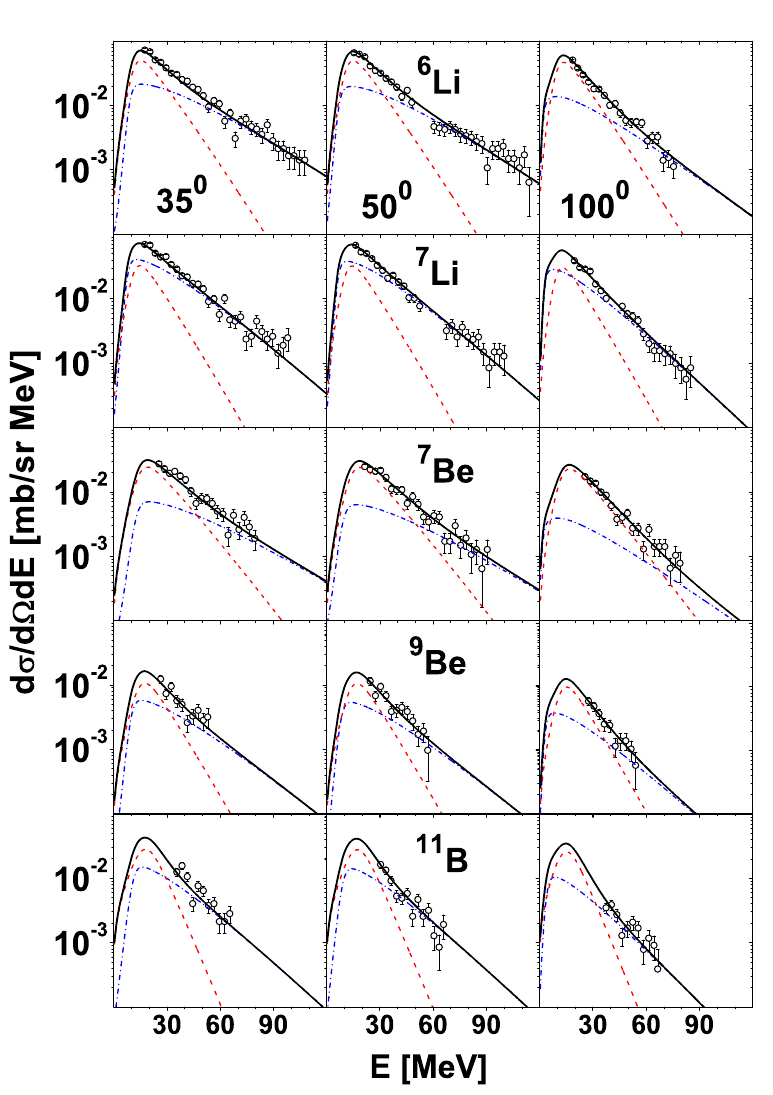}
\caption{\label{fig:libeb}  Typical spectra of lithium, beryllium,
and boron ejectiles from p+Ni collisions measured at 35$^{\circ}$,
50$^{\circ}$, and 100$^{\circ}$ (left, middle, and right columns,
respectively)  for 2.5 GeV proton beam impinging on to the Ni
target. The detected particles are listed in the central panel of
each row of pictures. Open circles represent the experimental data,
dashed, dot - dashed, and solid lines correspond to slow emitting
source, fast emitting source and the sum of both contributions,
respectively. }
\end{center}
\end{figure}
%
%
Two parameters were introduced to characterize the properties  of
the Coulomb barrier: $k$-parameter, i.e., height of the Coulomb
barrier in units of the height of the barrier B of two charged,
touching spheres of radius 1.44~A$^{1/3}$;
B~=~Z$_1$~Z$_2$~e$^2$/1.44~(A$_{1}^{1/3}$~+~A$_{2}^{1/3}$) MeV, and
the ratio B/d , where d is a diffuseness of the transmission
function through the barrier: $P(E)=(1+exp((E-kB)/d)))^{-1}$.
Details of this procedure, as well as interpretation of parameters
of the model can be found in Appendix of Ref. \cite{BUB07A}.

\subsection{\label{sec:IMF} Intermediate mass fragments}

The shape of the spectra of IMFs  is almost independent of the beam
energy. Thus, only the cross sections measured at 2.5 GeV beam
energy have been used for illustration of the quality of data
description. On the contrary, the shape of the spectra changes in
regular way with variation of the detection angle, namely the
spectra become more steep when the scattering angle increases. This
is especially well visible when the data are compared at angles
differing strongly as, e.g., 35$^{\circ}$ and 100$^{\circ}$.
Experimental spectra measured at three selected angles --
35$^{\circ}$, 80$^{\circ}$, and 100$^{\circ}$ -- are shown on Fig.
\ref{fig:libebinc} for most abundant isotopes of lithium, beryllium
and boron  as open circles together with the evaporation model
predictions (solid lines). Fluctuations of the theoretical spectra
are due to the Monte Carlo method of model calculations, i.e. due to
limited statistics of generated events. As can be seen, the
theoretical spectra are more steep than the experimental ones and
the absolute values of theoretical cross sections are several times
smaller than the data. The former effect is most pronounced for
$^{6,7}$Li and $^7$Be cross sections whereas the latter is present
 for all detected IMFs. Furthermore, the
evaporation model does not reproduce the tendency of shape variation
with the angle.

%
\begin{table*}
  \caption{\label{tab:parameters}Parameters of two moving sources  fitted to p+Ni data for isotopically identified
  IMF's and for $^4$He: $\beta_i$, $T_i$, and $\sigma_i$ correspond to  source velocity, its apparent temperature,
 and total production cross section, respectively. The sum $\sigma \equiv \sigma_1+\sigma_2$ is
 also listed.  The left part of the Table
 (parameters with indices "1")
 corresponds to the slow moving source, and the right part  contains values of parameters for the fast moving source.
  The upper row for each ejectile corresponds to beam energy 1.2 GeV, the row in the middle to 1.9 GeV,
  and the lowest one to the energy 2.5 GeV.
   }
\begin{tabular}{lrr|llr|cr}
  \hline \hline
            & \multicolumn{2}{c|}{Slow source} & \multicolumn{3}{c|}{Fast source}                  & \hspace*{0.5cm} $\sigma$  & \multicolumn{1}{r}{$\chi^2$} \\
  \cline{2-6}
  Ejectile   & \emph{T}$_1$/MeV & $\sigma_1$/mb & \hspace*{0.5cm} $\beta_2$   & \emph{T}$_2$/MeV & $\sigma_2$/mb & \hspace*{0.5cm} mb   &  \\
  \hline
  \hline
  $^4$He     &  7.0(2)          & 244(6)        & \hspace*{0.5cm} 0.040(3)    & 18.1(5)          &  76(6)        & \hspace*{0.5cm} 320(9)  & 7.3\\
             &  7.3(2)          & 269(6)        & \hspace*{0.5cm} 0.036(2)    & 19.4(5)          &  94(6)        & \hspace*{0.5cm} 363(9)  & 4.7\\
             &  7.9(2)          & 283(8)        & \hspace*{0.5cm} 0.035(2)    & 20.1(5)          & 101(7)        & \hspace*{0.5cm} 384(11) & 4.4\\
  \hline
  $^6$Li     &  9.1(4)          &  8.3(4)       & \hspace*{0.5cm} 0.035(3)    & 18.6(5)          &  4.1(5)       & \hspace*{0.5cm} 12.4(7)  & 1.5\\
             & 10.4(4)          & 11.5(6)       & \hspace*{0.5cm} 0.037(3)    & 19.8(5)          &  4.5(6)       & \hspace*{0.5cm} 16.0(9)  & 1.4\\
             &  9.4(5)          & 11.6(8)       & \hspace*{0.5cm} 0.026(3)    & 20.5(6)          &  8.0(9)       & \hspace*{0.5cm} 19.6(1.2)& 1.3\\
  \hline
  $^7$Li     &  8.1(7)          &  4.9(7)       & \hspace*{0.5cm} 0.022(2)    & 14.7(4)          &  6.6(9)       & \hspace*{0.5cm} 11.5(1.2)& 1.3\\
             &  9.6(7)          &  8.0(1.0)     & \hspace*{0.5cm} 0.025(3)    & 15.9(5)          &  7.0(1.2)     & \hspace*{0.5cm} 15.0(1.6)& 1.3\\
             &  9.6(1.0)        &  5.2(1.3)     & \hspace*{0.5cm} 0.018(2)    & 16.0(5)          & 12.3(1.8)     & \hspace*{0.5cm} 17.5(2.2)& 1.4\\
  \hline
  $^8$Li     & [8.0]            & [0.2]         & \hspace*{0.5cm} 0.040(8)    & 14.4(2.0)        &  1.2(5)       & \hspace*{0.5cm} 1.4(5)   & 1.4\\
             &  9.4(3.8)        &  0.6(5)       & \hspace*{0.5cm} 0.032(6)    & 17.2(1.1)        &  3.7(1.0)     & \hspace*{0.5cm} 4.3(1.1) & 1.0\\
             &  8.0(1.8)        &  1.3(4)       & \hspace*{0.5cm} 0.029(5)    & 18.0(1.0)        &  6.4(1.5)     & \hspace*{0.5cm} 7.7(1.6) & 1.0\\
  \hline
  $^7$Be     &  8.7(1.3)        &  2.8(5)       & \hspace*{0.5cm} 0.025(3)    & 16.8(7)          &  3.6(7)       & \hspace*{0.5cm} 6.4(9)   & 1.4\\
             &  9.5(9)          &  4.9(6)       & \hspace*{0.5cm} 0.025(3)    & 19.2(9)          &  3.9(8)       & \hspace*{0.5cm} 8.8(1.0) & 1.0\\
             & 11.3(7)          &  6.9(6)       & \hspace*{0.5cm} 0.032(5)    & 21.4(1.3)        &  2.6(7)       & \hspace*{0.5cm} 9.5(9)   & 1.0\\
  \hline
  $^9$Be     &  [8.6]           &  1.1(2)       & \hspace*{0.5cm} [0.023]     & 12.1(8)          &  1.3(2)       & \hspace*{0.5cm} 2.4(3)   & 1.1\\
             &  8.1(1.6)        &  2.0(5)       & \hspace*{0.5cm} 0.024(7)    & 14.2(1.2)        &  1.6(7)       & \hspace*{0.5cm} 3.6(9)   & 0.8\\
             &  8.3(2.0)        &  2.4(7)       & \hspace*{0.5cm} 0.019(7)    & 16.9(2.4)        &  1.8(1.1)     & \hspace*{0.5cm} 4.2(1.3) & 1.0\\
  \hline
  $^{10}$Be  & 6.1(1.9)         &  1.0(4)       & \hspace*{0.5cm} 0.030(16)   & 27.8(9.0)        &  0.4(2)       & \hspace*{0.5cm} 1.4(5)   & 1.8\\
             & 7.9(1.7)         &  1.0(4)       & \hspace*{0.5cm} [0.023]     & 17.6(3.1)        &  0.7(2)       & \hspace*{0.5cm} 1.7(5)   & 1.8\\
             & 6.1(1.9)         &  1.7(6)       & \hspace*{0.5cm} 0.024(11)   & 23.0(9)          &  0.8(4)       & \hspace*{0.5cm} 2.5(7)   & 1.3\\
  \hline
  $^{10}$B   & [6.0]            & 1.6(1.2)      & \hspace*{0.5cm} 0.018(4)    & 15.5(3.1)        &  1.7(6)       & \hspace*{0.5cm} 3.3(1.4) & 1.7\\
             & [6.0]            & 3.3(1.2)      & \hspace*{0.5cm} [0.023]     & 16.4(1.5)        &  2.4(4)       & \hspace*{0.5cm} 5.7(1.3) & 1.8\\
             & [6.0]            & 3.2(1.4)      & \hspace*{0.5cm} [0.023]     & 17.7(1.8)        &  2.7(4)       & \hspace*{0.5cm} 5.9(1.5) & 1.9\\
  \hline
  $^{11}$B   & [6.0]            & 3.1(9)        & \hspace*{0.5cm} [0.023]     & 13.4(1.2)        &  1.8(3)       & \hspace*{0.5cm} 4.9(1.0) & 1.3\\
             & [6.0]            & 5.2(1.2)      & \hspace*{0.5cm} [0.023]     & 17.8(2.2)        &  2.3(3)       & \hspace*{0.5cm} 7.5(1.3) & 1.5\\
             & [6.0]            & 5.9(3.6)      & \hspace*{0.5cm}  0.016(4)   & 14.8(3.2)        &  4.1(1.8)     & \hspace*{0.5cm} 10(4)    & 1.7\\
  \hline
  \hline
\end{tabular}
\end{table*}
%

All these facts indicate that an important contribution of another
reaction mechanism must be added to evaporation cross sections to
assure good reproduction of the data. This contribution is
comparable in magnitude with the evaporation cross section for Li
and becomes even more dominant for Be and B isotopes. Thus, the
IMF's data have been analyzed in the frame of a phenomenological
model of two moving sources as it was done for p+Au reactions
\cite{BUB07A}. In this way the  IMF's production on Ni target could
be compared with that for the Au target.

The parameters of two moving sources were searched for by fitting
the two moving sources cross sections  to experimental data, which
consisted of energy spectra measured at seven angles: 16$^{\circ}$,
20$^{\circ}$, 35$^{\circ}$, 50$^{\circ}$, 65$^{\circ}$,
80$^{\circ}$, and 100$^{\circ}$. To decrease the number of
parameters it was assumed that the velocity of the slow source
emitting IMF's is equal to the velocity of the heavy residuum from
intranuclear cascade, i.e., $\beta_{1}$=0.005. The mean values of
this velocity was found in calculations of intranuclear cascade to
be equal to 0.0051c, 0.0049c, and 0.0047c for 1.2, 1.9, and 2.5 GeV
beam energy, respectively. It was checked that the modification of
this parameter by 30\% causes changes of other parameters smaller
than their errors estimated by fitting computer program. In
evaluation of $k$-parameter it was assumed that $B$ - value is
defined as the Coulomb barrier between the emitted particles and the
target nucleus.  The results of the fit are not very sensitive to
the value of the $k$-parameter because the experimental low energy
limit of the spectra is above the position of the Coulomb barrier
for most of IMFs .  Thus fixed values of $k_1=0.75$ and $k_2=0.3$
were used. The $B/d$ ratio was arbitrarily assumed to be equal to
5.5.

 Values of the best fit parameters are
listed in  Table \ref{tab:parameters}. The errors of the parameters
are also given  when the program searching for the best fit could
estimate them. However, sometimes program was not able to estimate
the errors, especially when strong ambiguities of parameters were
present. In such a case the values of the parameters are quoted
without estimation of errors. Sometimes it was useful to fix the
values of parameters to avoid numerical problems leading to
ambiguities. Such values are quoted in the table in parentheses.

A very good description of the spectra of all IMF's has been
obtained as can be judged from the inspection of Fig.
\ref{fig:libeb} and values of the $\chi^{2}$  given in the Table
\ref{tab:parameters}, which usually vary between 1 and 2.

It is obvious from comparison of Figs. \ref{fig:libebinc} and
\ref{fig:libeb} that the spectra of particles emitted from the slow
source are very similar  to the evaporation spectra of particles
from the residuum of intranuclear cascade, however, it is not the
case for the spectra of the fast source.  They have much smaller
slope and are dependent on the scattering angle.  It is evident that
their presence is necessary to reproduce the high energy part of
experimental spectra.

\begin{figure}
\begin{center}
\vspace*{-0.5cm}
\includegraphics[angle=0,width=0.5\textwidth]{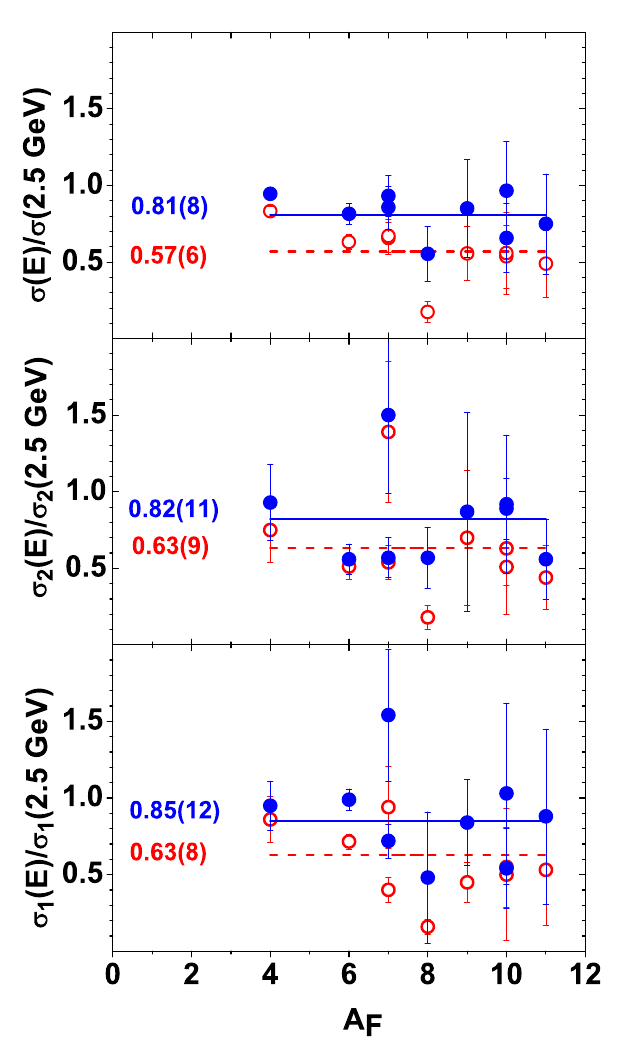}
\caption{\label{fig:sov2p5L}  Ratio of production cross section
determined at 1.2 (open circles) and 1.9 GeV (full dots) to the
cross sections found at 2.5 GeV beam energy. The lower panel
presents cross sections for slow sources, panel in the middle for
fast sources, and upper panel for sum of both contributions. The
horizontal lines depict values of the ratios averaged over IMFs; the
solid lines for the 1.9 GeV , and the dashed lines for 1.2 GeV. }
\end{center}
\end{figure}

The cross sections of both emitting sources increase with beam
energy.  This is illustrated on Fig.\ref{fig:sov2p5L} where ratios
of the production cross sections for beam energy 1.2 and 1.9 GeV to
the cross sections for 2.5 GeV are presented.  It turned out that
the increase of the cross sections with the beam energy is the same,
in the limits of errors, for contributions emerging from both
emitting sources.  The cross sections increase approximately by
factor 1.7 when beam energy increases from 1.2 GeV to 2.5 GeV (cf.
numbers depicted in Fig. \ref{fig:sov2p5L}).  Since the energy
dependence is very similar for both emitting sources, the relative
contributions of the sources remain constant in the studied energy
range and, moreover, both are equal in the limits of errors: The
averaged over IMFs ratio of $\sigma_2/(\sigma_1 + \sigma_2)$ is
equal to 0.50(14), 0.46(11), and 0.48(16) for beam energy 1.2, 1.9,
and 2.5 GeV, respectively. Thus, this ratio averaged over three
energies is equal to 0.48(7).

The parameters of the sources, which influence the shape of the
spectra (velocity $\beta$ and temperature $T$) should not change
with the beam energy because the shape of experimental spectra is
independent of the beam energy (cf. Fig. \ref{fig:helibeb}).
However, there is a distinct difference for each energy between
values of the parameters characterizing the slow source and the fast
source. The velocity of the fast source $\beta_2$, averaged over
IMFs is equal to 0.027(3), 0.027(2), and 0.023(2) for beam energy
1.2, 1.9, and 2.5 GeV, respectively. These values are about five
times larger than fixed velocity of the slow source ($\beta_1 =
0.005$). The temperature parameter $T_2$ of the fast source  is
about two times larger than the temperature parameter of the slow
source $T_1$. Its values (averaged over IMFs)  are equal to
16.8(1.2), 17.4(6), and 17.8(8) MeV for the fast source and  7.6(3),
8.4(6), and 8.1(5) MeV for the slow source at three studied beam
energies 1.2, 1.9, and 2.5 GeV, respectively.

\begin{figure}
\begin{center}
\vspace*{-0.5cm} \hspace*{-0.5cm}
\includegraphics[angle=0,width=0.5\textwidth]{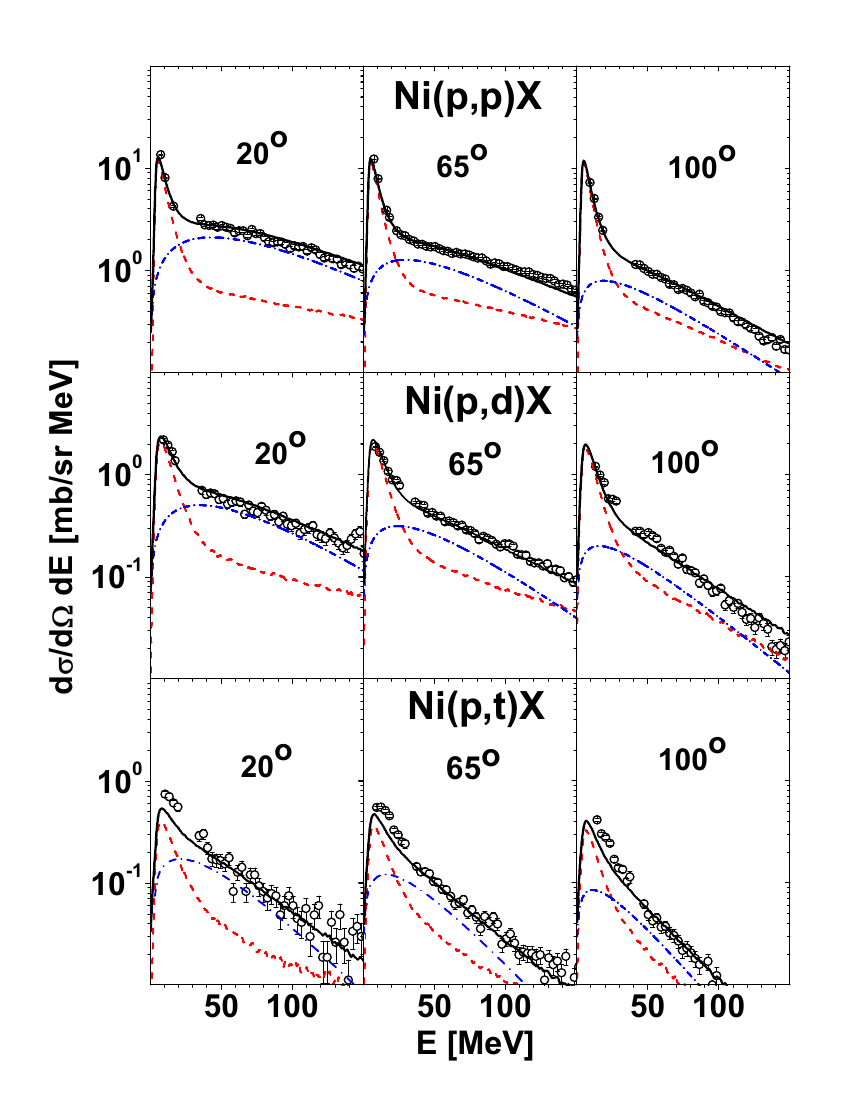}
\caption{\label{fig:pdt}  Typical spectra of protons, deuterons, and
tritons (upper, middle, and lower rows of the figure, respectively)
measured at 20$^{\circ}$, 65$^{\circ}$, and 100$^{\circ}$ (left,
middle, and right columns of the figure, respectively)  for 2.5 GeV
proton beam impinging on to the Ni target. Open circles represent
the experimental data, dashed, dot - dashed, and solid lines
correspond to the two-step model, the emission from the fireball and
the sum of both contributions, respectively. }
\end{center}
\end{figure}

\subsection{\label{sec:LCP} Light charged particles}

\begin{figure}[H]
\begin{center}
\vspace*{-0.5cm} \hspace*{-0.5cm}
\includegraphics[angle=0,width=0.44\textwidth]{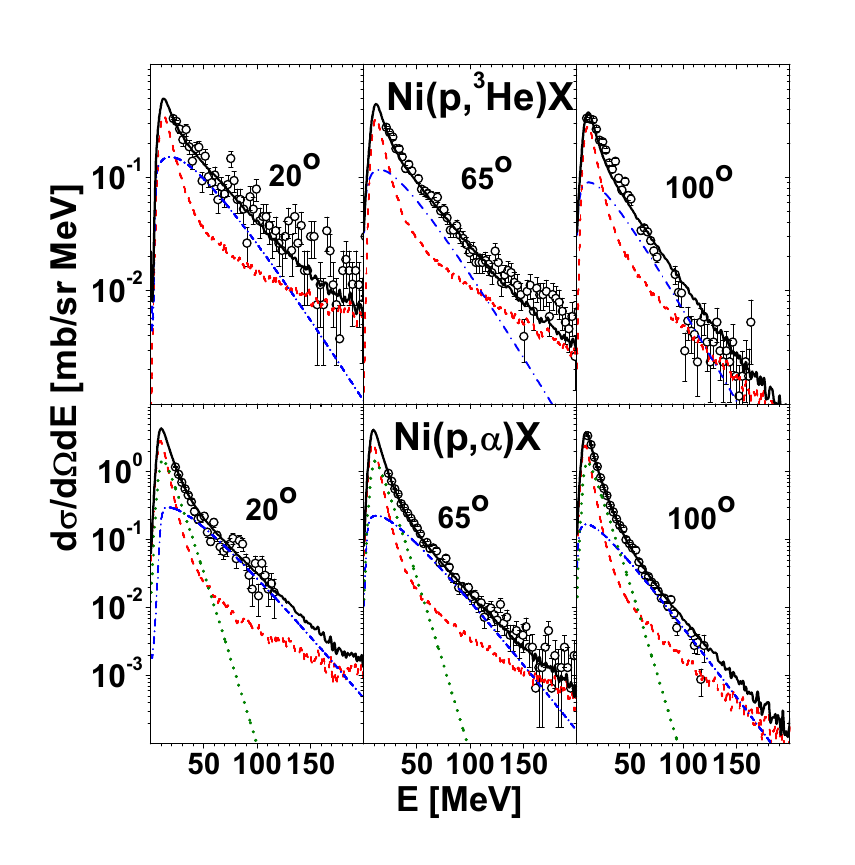}
\caption{\label{fig:he34}  
Same as on Fig. \ref{fig:pdt} but for $^3$He and $^4$He. 
The dotted line for $^4$He denotes the contribution of an
additional, slowly moving source. }
\end{center}
\end{figure}

All experimental spectra of LCPs from p+Ni collisions contain large
high energy component (cf. Figs \ref{fig:helibeb}, \ref{fig:pdt},
and \ref{fig:he34}), which cannot be reproduced by evaporation of
particles from the equilibrated remnant of the intranuclear cascade.
Thus, the nonequilibrium emission of LCPs must play an important
role. The coalescence of nucleons escaping from the target nucleus
together with nucleons taking part in the intranuclear cascade may
lead to such an emission.  Indeed,  Boudard et al. \cite{BOU04A} and
Letourneau et al. \cite{LET02A} have shown that the microscopic
calculation of coalescence occuring when the intranuclear cascade
proceeds is able to reproduce a large part of the observed effect.
However, it was found that the improvement of the description of
LCPs spectra deteriorates simultaneously the proton spectra because
increasing of the production of composite particles occurs on the
account of decreasing emission of the nucleons.  This contradiction
led the present authors to search for another nonequilibrium
process, which could be responsible for the observed enhancement of
the high energy part of the LCP spectra.  It was proposed
\cite{BUD08A}, that the emission from a fireball, i.e., a fast and
hot group of nucleons consisting of target nucleons lying on the
straight way of the bombarding proton through the target nucleus
\cite{WES76A}, can account for the missing nonequilibrium component
of the LCP cross sections. A sum of the coalescence of nucleons and
the emission of LCPs from the fireball was found to be crucial for
the very good description of the data for proton induced reactions
on Au target at three beam energies: 1.2, 1.9, and 2.5 GeV
\cite{BUD08A}. Furthermore, the emission of the fireball, which
introduces a strong rearrangement of the target nucleus can lead to
a break-up of the remnant of the target and thus to an appearing of
two moving sources also emitting intermediate mass fragments and
LCps. Therefore, this hypothesis explains simultaneously the
presence of the nonequilibrium emission for IMFs which was discussed
above.

Since the fireball contains only several nucleons its contribution
is present only for LCPs.  On the contrary, the fast and slow
excited prefragments of the target may emit IMFs as well as LCPs. In
the present analysis their contribution to spectra of LCPs has been,
however, neglected because it was estimated (by extrapolation of
parameters found for IMFs  to lighter ejectiles) to be much smaller
than contributions of other reaction mechanisms. Magnitude of cross
sections for emission of LCPs from two sources - products of
break-up -- were estimated to be $\sim$ 10\% of the total cross
sections, i.e., to be of order of errors of the fitting procedure.

In the present study the same procedure of the description of LCP
spectra as that in ref. \cite{BUD08A} has been applied. The INCL4.3
computer program \cite{BOU04A} has been used for the description of
the intranuclear cascade of nucleon-nucleon collisions with
inclusion of coalescence of nucleons, whereas the GEM2 computer
program \cite{FUR00A,FUR02A} served for evaluation of evaporation of
particles from heavy target residuum remaining after the
intranuclear cascade.  The default parameter values, proposed by the
authors of both programs, have been used, respectively.

Since there is no explicit room for the presence of the fireball in
the microscopic calculations performed according to the intranuclear
cascade model, the inclusion of fireball emission should be
accompanied by decreasing the contribution from direct processes
simulated by intranuclear cascade and coalescence of escaping
nucleons. Thus, the spectra of protons evaluated from intranuclear
cascade with inclusion of coalescence and with contribution of
evaporation of particles were multiplied by a factor F, common for
all scattering angles, treated as a free parameter and then added to
the contribution from the fireball emission calculated according to
the formula of single moving source emitting isotropically the LCP's
\cite{WES78A}. The same, fixed value of the scaling factor has been
used for further analysis of data for other LCPs.  The parameters of
the single moving source - the fireball, i.e. its temperature
parameter - $T_3$, velocity of the source - $\beta_3$, total
production cross section associated with this mechanism - $\sigma_3$
were treated also as free parameters.

\begin{table*}
 \caption{\label{tab:fireball} Parameters of the fireball  fitted to p+Ni data; $\beta_3$, $T_3$,
 and $\sigma_3$ correspond to fireball velocity in units of speed of light, its apparent temperature,
 and total production cross section, respectively.
 The upper row for each ejectile corresponds to beam energy 1.2 GeV, the row in the middle to 1.9 GeV,
  and the lowest one to the energy 2.5 GeV.
 Parameter F is the scaling factor of coalescence and evaporation
 contribution extracted from fit to the proton spectra. The numbers in parentheses
 show fixed values of the parameters. Note, that for the $\alpha$ particles contribution of two additional
 moving sources should be added with parameters given in  Table
 \ref{tab:parameters}. The columns described as F$ \ast \sigma_{INCL}$ and F$\ast \sigma_{GEM}$ contain total
 production cross sections due to intranuclear cascade with the coalescence and due to evaporation
 from the target residuum, respectively. The total production cross section obtained by summing of all contributions
 is depicted in the column denoted by $\sigma$. In the case of alpha particles it contains also
 the contribution of the emission from slow and fast sources listed in Table \ref{tab:parameters}.}

\begin{center}
 \begin{tabular}{lcrrccrrr}
\hline \hline
  Ejectile & $\beta_3$  &  $T_3$    &  $\sigma_3$ &     F   & F$\ast \sigma_{INCL}$ & F$\ast\sigma_{GEM}$ & \hspace*{0.5cm} $\sigma$ & $\chi^{2} $ \\
           &            &  MeV      &  mb         &         &    mb             &   mb             & \hspace*{0.5cm} mb      & \\
\hline
  \emph{p}  & 0.149(12) & 38.9(2.1) & 1071(61)    & 0.70(3) &  1094             &    994           & \hspace*{0.5cm} 3159(61) & 179 \\
            & 0.156(10) & 41.7(1.9) & 1222(53)    & 0.70(2) &  1139             &   1005           & \hspace*{0.5cm} 3366(53) & 95.1 \\
            & 0.163(8)  & 43.2(1.5) & 1343(44)    & 0.79(2) &  1286             &   1123           & \hspace*{0.5cm} 3752(44) & 48.8 \\
 \hline
  \emph{d}  & 0.105(4)  & 32.5(8)   &  181(5)     & [0.70] &    202             &    174           & \hspace*{0.5cm} 557(5)   & 9.5   \\
            & 0.099(3)  & 33.7(6)   &  234(5)     & [0.70] &    201             &    196           & \hspace*{0.5cm} 631(5)   & 3.2 \\
            & 0.100(4)  & 35.8(8)   &  272(7)     & [0.79] &    220             &    225           & \hspace*{0.5cm} 717(7)   & 7.4 \\
\hline
  \emph{t}  & 0.062(3)  & 21.9(6)   &  41.9(1.6)  & [0.70] &    40.9            &    28.9          & \hspace*{0.5cm} 111.7(1.6) & 1.3 \\
            & 0.055(3)  & 23.8(6)   &  59.8(2.1)  & [0.70] &    41.3            &    34.1          & \hspace*{0.5cm} 135.2(2.1) & 1.4 \\
            & 0.054(3)  & 25.0(6)   &  71.5(2.2)  & [0.79] &    45.3            &    39.4          & \hspace*{0.5cm} 156.2(2.2) & 1.3 \\
\hline
 $^{3}$He   & 0.046(2)  & 22.9(5)   &  43.1(1)    & [0.70] &    31.2            &    32.6          & \hspace*{0.5cm} 106.6(1)   & 3.4 \\
            & 0.039(2)  & 23.5(4)   &  60.5(1.2)  & [0.70] &    31.6            &    37.5          & \hspace*{0.5cm} 129.6(1.2) & 3.3 \\
            & 0.040(2)  & 25.0(5)   &  69.4(1.4)  & [0.79] &    34.5            &    43.1          & \hspace*{0.5cm} 147.0(1.4) & 2.9 \\
\hline
 $^{4}$He  &            &           &             & [0.70] &    16.8            &    277           & \hspace*{0.5cm} 614(9)     & 7.3 \\
           &            &           &             & [0.70] &    16.6            &    281           & \hspace*{0.5cm} 661(9)     & 4.7 \\
           &            &           &             & [0.79] &    18.1            &    312           & \hspace*{0.5cm} 714(11)    & 4.4 \\
\hline \hline
\end{tabular}
\end{center}
 \end{table*}
%

Parameters $k_3$ (the height of the Coulomb barriers in units of $B$
- Coulomb barrier between the ejectile and the target nucleus) and
parameter $B/d$ describing diffuseness of the transmission function
through the Coulomb barrier are fixed at arbitrarily assumed values
0.07 and 4.8, respectively.  Values of the fitted parameters are
collected in the Table \ref{tab:fireball}.

The fit was performed for 7 scattering angles (16$^{\circ}$,
20$^{\circ}$, 35$^{\circ}$, 50$^{\circ}$, 65$^{\circ}$,
80$^{\circ}$, and 100$^{\circ}$). Results of the fit are presented
in Fig. \ref{fig:pdt}  for protons, deuterons and tritons, and in
Fig. \ref{fig:he34} for $^3$He and $^4H$e.  Since the spectra at
various beam energies almost do not differ in the shape, the
comparison of theoretical cross sections with the data is shown only
for one beam energy, namely for 2.5 GeV.  The left column on both
figures represents cross sections for 20$^0$, the column in the
middle for 65$^0$, and the right column the data measured at
100$^0$.  The proton spectra are shown in the upper row of the Fig.
\ref{fig:pdt}, the deuteron and triton data in the middle and lower
row, respectively. The $^3$He data are depicted in the upper row of
Fig. \ref{fig:he34}, whereas the $^4$He cross sections occupy the
lower row of this Figure.

It is obvious from inspection of Figs. \ref{fig:pdt},
\ref{fig:he34}, that a very good description of  the experimental
data for all LCPs was achieved.  It should be emphasized, that the
values of the best fit parameters vary smoothly from ejectile to
ejectile as well as from the one beam energy to another, thus the
same mechanism seems to be responsible for the nonequilibrium
processes for all these particles. It was, however, found that
values of the parameters, necessary
%
%
%
\noindent to describe the alpha particle data differ from those for
lighter LCPs. It turned out that (i) it is necessary to use two
emitting sources instead one fireball for good data reproduction,
and (ii) the parameters of these sources have quite similar values
as those for IMFs (cf. Table \ref{tab:parameters}). For this reason,
it may be concluded that the alpha particles behave rather like IMFs
than as LCPs.

\begin{figure}[H]
\begin{center}
\includegraphics[angle=0,width=0.5\textwidth]{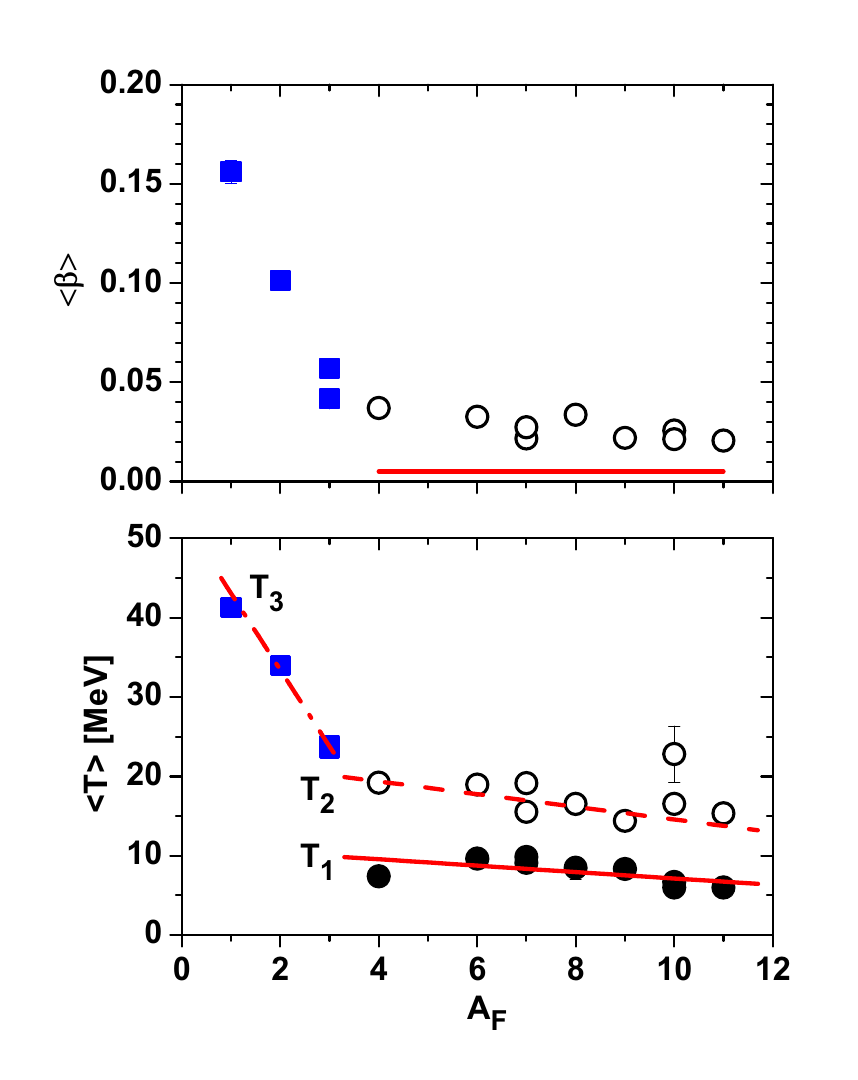}
\caption{\label{fig:bt123L}  In the lower panel of the figure the
apparent temperature of the moving sources, averaged over beam
energies is drawn as a function of the ejectile mass. Open circles
and full dots represent values of temperature parameters $T_2$ and
$T_1$ for fast and slow source, respectively. Full squares indicate
temperature $T_3$ of the fireball.  The solid and dashed lines were
fitted to the points representing the IMF's and $^4$He. Dash dotted
line was fitted to points representing the LCP's. In the upper panel
of the figure the dependence of the beam energy averaged velocity of
the sources is drawn versus mass of ejectiles. The symbols and lines
have the same meaning as for the lower part of the figure with one
exception: The full dots are not shown because the velocity of
slower source was fixed
 during analysis (at velocity $\beta_1$=0.005 of heavy residuum of
target nucleus after intranuclear cascade) and it is represented by
solid line in the figure. }
\end{center}
\end{figure}

Values of fireball velocity $\beta_3$ and its temperature parameter
$T_3$ do not change systematically with the beam energy and their
fluctuations are so small, that it is possible to assume that they
do not change with the energy.  Similar situation was observed for
IMFs. Thus the energy averaged values of velocities and temperature
parameters of all sources are collected in one figure - Fig.
\ref{fig:bt123L} - to allow for discussion of their dependence on
the mass of ejectiles.
%
%

The velocities and temperature parameters are grouped in three
distinctly separated sets corresponding to the slow source
($\beta_1$ and $T_1$), to the fast source ($\beta_2$ and $T_2$), and
to the fireball ($\beta_3$ and $T_3$). The mass dependence -
approximated by straight lines - is also different for each source.
The slope of the mass dependence is the smallest for the slow
source, of intermediate value for the fast source, and the largest
for the fireball.  The linear dependence of the temperature
parameter on the mass of ejectiles is expected as a result of
momentum conservation, i.e., the recoil of the source of a given
mass appears during the emission of ejectiles of various masses.
This linear dependence of the temperature parameter on the mass of
ejectile allows for the extraction of recoil corrected temperature
of the source - $\tau$ and the estimation of the mass of the source
A$_S$. If the mass of the source is constant - the same for all
ejectiles - the determination of its mass and recoil corrected
temperature is unambiguous. However, there are arguments that the
source must have some distribution of masses with different average
value for each ejectile. For example, the deuterons cannot be
emitted by fireball consisted of only two nucleons but such emission
may occur from the fireball built of three nucleons. The emission of
protons, on the contrary, can appear both from two-nucleon and
three-nucleon source, thus the mass of the fireball emitting protons
is in average smaller than the mass of fireball emitting the
deuterons.  This may be a reason of strong decreasing of the
fireball velocity with the mass of the ejectile as well as only
slight decreasing of the velocity of the fast source with the mass
of IMFs.  A change of the mass of the fireball by one nucleon is
very significant, because the fireball may be built of only several
nucleons, whereas such a change for the source consisted of 20 or
more nucleons is hardly to be observed in the mass dependence of the
temperature parameter.

The given above arguments show that the extraction of the mass of
the source and its recoil corrected temperature from the ejectile
mass dependence of the temperature parameter should be taken with
caution and treated only as a crude estimation.  Such an estimation
is discussed below and extracted parameters are compared with those,
which were found in our previous study \cite{BUD08A} of reactions
induced by protons on the gold target.


\section{\label{sec:discussion}Discussion}

The parameters of linear functions describing the dependence on the
ejectile mass of the  temperature parameter $T$ and velocity $\beta$
of three sources  are collected in the Table \ref{tab:AuNi}. Those
parameters, obtained in the previous study of reactions induced by
protons on the gold target \cite{BUD08A}, are also listed in this
Table.

\begin{table}
\caption{\label{tab:AuNi} Beam energy averaged temperature and
velocity parameters of three sources of ejectiles for Au and Ni
targets. $T$ denotes apparent source temperature (in MeV), $\tau$ -
the temperature parameter corrected for the recoil, $A_S$ represents
mass number of the source, and $\beta$ its velocity in units of
speed of light. The symbol $A$ indicates the mass number of the
ejectile. Parameters with index $1$ correspond to slow source, with
index $2$ to fast source, and with index $3$ to the fireball. }
\begin{center}
 \begin{tabular}{c|c|c}
\hline \hline
   Parameter     &    Au                   &    Ni    \\
\hline
 $T_1$           &   11.1(3)               & 11.2(7) - 0.4(2)$\ast$A \\
 $\tau_1$        &   11.1(3)               & 11.2(7)   \\
 A$_{S_1}$       &   $\sim$ 165                 &  28(15)  \\
 $\beta_1$       &  [0.003]                &  [0.005] \\
\hline
 $T_2$           & 30.6(4) - 1.61(45)$\ast$A    & 22.5(6) - 0.8(1)$\ast$A \\
 $\tau_2$        & 30.6(4)                 & 22.5(6) \\
 A$_{S_2}$       &   19(6)                 & 28(4) \\
 $\beta_2$         & 0.059(5) - 0.0034(6)$\ast$A  & 0.044(6) - 0.0021(7)$\ast$A \\
\hline
 $T_3$           & 49.9(7) - 8.2(2.6)$\ast$A    & 52.7(1.1) - 9.6(4)$\ast$A \\
 $\tau_3$        & 49.9(7)                 & 52.7(1.1) \\
 A$_{S_3}$       &    6(2)                 & 5.5(3) \\
 $\beta_3$         & 0.218(39) - 0.051(16)$\ast$A & 0.209(11) - 0.053(5)$\ast$A   \\
 \hline
\end{tabular}
\end{center}
\end{table}

It is clearly visible that\emph{ all properties of the fireballs for
both targets are identical in the limits of errors}. This seems to
be unexpected, especially as concerns masses of both fireballs,
because of large difference between the mass of Ni and Au targets.
However, according to the simple picture of the fireball, it is
consisted of the nucleons which lie on his straight way through the
target nucleus, therefore the mass of the fireball should scale as
$A^{1/3}$.  It means that the ratio of fireball masses for Au and Ni
should be equal to $\sim 1.5$. This ratio, extracted from the
phenomenological analysis, is equal to 1.1(7) what means, that  in
the limits of errors it is in agreement with the assumed picture of
the mechanism.

The equality of velocities of fireballs and their temperatures for
both targets may be treated as consequence of the same momentum and
energy transfer from the bombarding proton to the group of nucleons
forming the fireball.  Such an explanation is in line with the fact
of the same beam energies for both targets and small difference in
the thickness of the nuclear matter placed on the way of bombarding
proton.  Therefore, this equality may be interpreted as the argument
in favor of assumed model of the reaction.

The recoil corrected temperatures of slow sources for Ni and Au
targets are also the same.  Of course, the mass $A_{S_2}$ of the
slow source is completely different in the case of Au target ($\sim$
165) and Ni target ($\sim$ 28 ). Therefore the recoil correction of
the temperature parameter could be neglected for the Au target but
is visible in the ejectile mass dependence of the temperature
parameter $T_1$ for the Ni target.  This difference of source masses
reflects also on the velocity of the slow source $\beta_1$. 

The largest differences appear for the fast source.  It should be,
however, pointed out that the parameters do not differ in average
more than $\sim$ 50\%. Taking into consideration the fact of limited
accuracy of extraction of values of the parameters it may be claimed
that the parameters of the fast source are similar for both targets.

The above considerations concern \emph{velocity and temperature of
three sources for Au and Ni targets} and show, that this parameters,
which are almost independent of the beam energy in the studied
proton energy range 1.2 -- 2.5 GeV, \emph{are very similar for both
targets.}

In the following the behavior of production cross sections will be
discussed.  As it was shown on Fig. \ref{fig:sov2p5L}, cross
sections for all IMFs increase in average by factor $\sim$ 1.7 when
proton beam energy increases from 1.2 GeV to 2.5 GeV.  This is true
for the total production cross sections as well as for the
contributions of individual emitting sources, however, the spread
(among various IMFs) of the ratio of given cross section to that
measured at 2.5 GeV is smaller for total production cross section
than for individual sources.  The ratios of the cross sections
measured at 1.2 and 1.9 GeV to those measured at 2.5 GeV are listed
in Table \ref{tab:sigratio}.

\begin{table}
\caption{\label{tab:sigratio} Averaged over IMFs ratio of cross
section $\sigma_i$ for energy $E$ to the same cross section measured
at proton beam energy 2.5 GeV  on Au and Ni targets. The $\sigma_i$
is equal to $\sigma_1$ (slow source), $\sigma_2$ (fast source) or
$\sigma \equiv \sigma_1+\sigma_2$ (total production cross section) }
\begin{center}
 \begin{tabular}{c|ccc}
\hline \hline
 Ratio of $\sigma_i$ to $\sigma_i$(2.5 GeV) & at E/GeV   &    Au      &   \hspace*{0.5cm} Ni    \\
\hline
 $\sigma_1$               & 1.2              &   0.39(3)  & \hspace*{0.5cm} 0.63(8)  \\
                          & 1.9              &   0.75(6)  & \hspace*{0.5cm} 0.85(12) \\
\hline
 $\sigma_2$               & 1.2              &   0.23(3)  & \hspace*{0.5cm} 0.63(9)  \\
                          & 1.9              &   0.57(6)  & \hspace*{0.5cm} 0.82(11)  \\
\hline
 $\sigma_1+\sigma_2$      & 1.2              &   0.33(2)  & \hspace*{0.5cm} 0.57(6)  \\
                          & 1.9              &   0.66(4)  & \hspace*{0.5cm} 0.81(8)  \\
\hline
\end{tabular}
\end{center}
\end{table}

\begin{figure}
\begin{center}
\hspace*{-1.0cm}
\includegraphics[angle=0,width=0.5\textwidth]{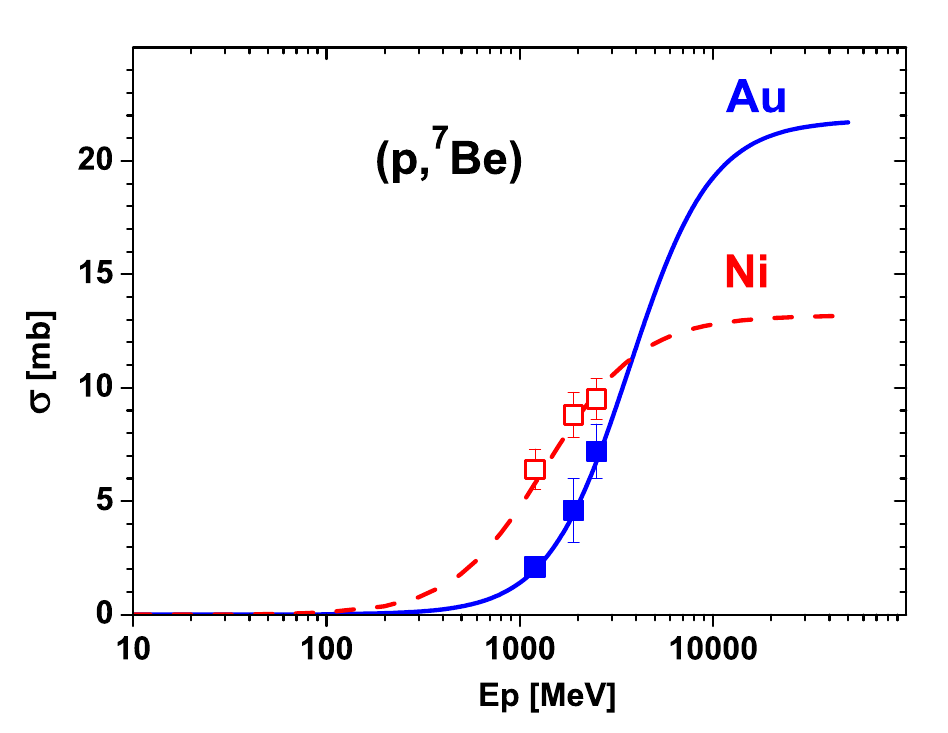}
\caption{\label{fig:be7endep}  Energy dependence of production cross
section of $^7$Be ejectiles in proton induced reactions.  The lines
show results of compilation of $^7$Be production cross sections
\cite{BUB04A}, the symbols represent experimental data of the
present experiment (open squares) for Ni target and the data
published by Budzanowski \emph{et al.} \cite{BUD08A} (full squares)
for Au target.  The solid and dashed lines depict the excitation
functions from Ref. \cite{BUB04A} for Au and Ni targets,
respectively. }
\end{center}
\end{figure}

It is clear from examination of Table \ref{tab:sigratio} that the
IMF production cross sections measured for Au target increase
stronger with beam energy than those for Ni target.  This can be
understood as an effect caused by difference between threshold
energies for fragmentation of both targets. To illustrate this
effect the excitation function for $^7$Be production is shown on
Fig. \ref{fig:be7endep}  as a typical example.  It is seen that
fragmentation starts at lower energies on Ni target than on Au
target.  Therefore the studied range of beam energy (1.2 - 2.5 GeV)
corresponds for Ni target to the region where the production cross
section starts to saturate, whereas for Au target this is region
where the production cross section starts to rise quickly.
Furthermore, the leveling of the production cross section for the Ni
target appears at lower value than that for Au target. Both these
effects cause that the total production cross sections should rise
more quickly for the Au target than for Ni target, in accordance
with present observations.

\begin{figure}
\begin{center}
\hspace*{-1.0cm}
\includegraphics[angle=0,width=0.55\textwidth]{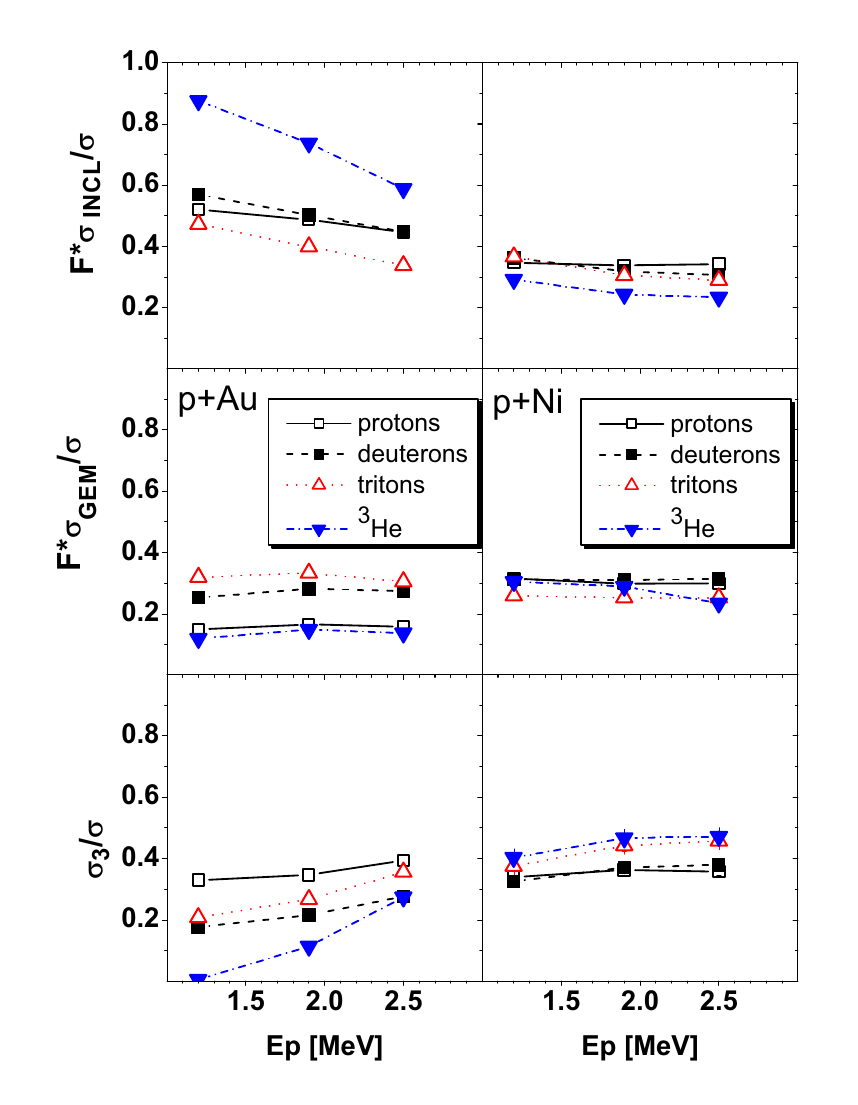}
\caption{\label{fig:figcontrL}  Energy dependence of various
reaction mechanisms for protons (open squares), deuterons (full
squares), tritons (open triangles) and $^3$He (full triangles).  The
relative contribution of the fast stage of the reaction, i.e.,
intranuclear cascade and coalescence of nucleons into LCPs
$\sigma_{INCL}/\sigma$ is presented in the upper panel of the
figure, evaporation contribution from the equilibrated residuum of
the intranuclear cascade $\sigma_{GEM}/\sigma$ is shown in the
middle panel, whereas the contribution of fireball emission
$\sigma_3/\sigma$ is depicted in the lower panel. }
\end{center}
\end{figure}

It is interesting to examine how large are the contributions of
individual reaction mechanisms for emission of LCPs, i.e.,  the
preequilibrium stage of the reaction described by intranuclear
cascade and coalescence of nucleons into composite particles, the
fireball emission, and the evaporation.
 As it is seen in Fig. \ref{fig:figcontrL} (right column)
the contribution from intranuclear cascade and coalescence is for Ni
target almost equal to that from evaporation and exhausts about 30\%
of the total production cross section for all studied beam energies
and ejectiles. The first of these contributions decreases several
percent with the energy, whereas the evaporation contribution is
almost independent of the energy. The contribution of the fireball
is slightly larger ($\sim$ 40 \%) and increases several percent in
the studied energy range. These variations are more pronounced for
tritons and $^3$He than for protons and deuterons.

This picture is quite different from energy behavior of separate
reaction mechanisms of LCPs production for the Au target (cf. Fig.
\ref{fig:figcontrL}, left column), with exception of the evaporation
contribution ($\sigma_{GEM}$ on the figure) which is almost energy
independent similarly as for Ni target.  The coalescence
contribution decreases with beam energy ($\sim$ 10\% for p,d,t, and
$\sim$ 30\% for $^3$He) but fireball contribution increases (also
$\sim$ 10\% for p,d,t, and even more for $^3$He).  Moreover, the
coalescence contribution for Au target is in average larger,  and
the fireball contribution smaller than for Ni target. Such a
behavior of relative contributions of various processes might
suggest that they depend rather on the proton beam energy per
nucleon of the target than on the beam energy itself, similarly as
it is in the case of the total production cross section (cf. Figs.
\ref{fig:be7endep} and \ref{fig:figcontrL}).

\section{Summary and conclusions}

A new, extensive set of double differential cross sections
$d^{2}\sigma/d\Omega dE$ for the production of LCPs and light IMFs
($^{6}$He, $^{6,7,8}$Li, $^{7,9,10}$Be, $^{10,11}$B,and C) in
collision of protons with Ni target has been measured at three beam
energies (1.2, 1.9, and 2.5 GeV).  The data were analyzed  using
two-step microscopic model - intranuclear cascade with possibility
to take into account coalescence of nucleons into LCPs, followed by
evaporation of particles from the equilibrated residual nuclei.
Since the model could not satisfactorily reproduce the spectra
measured for seven scattering angles; 16$^0$, 20$^0$, 35$^0$,
50$^0$, 65$^0$, 80$^0$, and 100$^0$, the phenomenological analysis
has been performed assuming the emission of LCPs from the fireball,
i.e., fast and hot source built of several nucleons moving in the
forward direction, as well as emission of IMF from two slower and
colder sources, which were interpreted as prefragments of the target
appearing due to break-up of the target.  An excellent description
of all data has been achieved with smoothly varying values of the
parameters from ejectile to ejectile.  Due to such good reproduction
of energy and angular dependencies of $d^{2}\sigma/d\Omega dE$ by
model calculation it was possible to determine total production
cross sections for all studied ejectiles.  They are listed in Table
\ref{tab:parameters} and \ref{tab:fireball} for IMFs and LCPs,
respectively.

It turned out that the competition of the emission from the fireball
with the coalescence process and evaporation of particles from the
equilibrated remnant of intranuclear cascade is crucial for a proper
description of the data for LCPs.  The emission of IMFs from a fast
moving source was necessary to reproduce the high energy part of the
experimental spectra.  Moreover, the emission of IMFs from a slow
moving source, which emits particles with spectra similar to those
of evaporated IMFs from the equilibrated residua of intranuclear
cascade, improved significantly the description of the low energy
part of the spectra.

In summary, a large contribution of the nonequilibrium processes,
which are not contained in the two-step microscopic model, has been
established.  Properties of these processes are compatible with
hypothesis of emission of ejectiles from three moving sources: The
light, fast, and hot source - a fireball, which appears as result of
knock-out of group of nucleons lying on the straight way of the
impinging proton through the nucleus, and two slower and colder
sources, which are created due to break-up of the target remnant
after emission of the fireball.

All the discussed above effects are very similar to those observed
by present authors for another nuclear system, namely p+Au in the
same range of beam energies \cite{BUB07A,BUD08A}.  As it was
mentioned in the Introduction, the Ni target is significantly
different from Au target.  Thus the observation of analogous
reaction mechanisms in both nuclear systems suggests that this
mechanism appears generally. Properties of this mechanism are in
many aspects the same for Ni and for Au targets but some clear
differences are also present. Therefore further investigations
should be performed  for other nuclei and beam energies to achieve
reasonable progress toward the understanding of the fragmentation
phenomenon, and to allow for good description of the production
cross sections of LCPs and IMFs.

\begin{acknowledgments}

The technical support of A.Heczko, W. Migda{\l}, and N. Paul in
preparation of experimental apparatus is greatly appreciated. This
work was supported by the European Commission through European
Community-Research Infrastructure Activity under FP6 project Hadron
Physics, contract number RII3-CT-2004-506078). This work was also
partially supported by the Helmholtz Association through funds
provided to the virtual institute "Spin and strong QCD" (VH-VI-231).
One of us (M.F.) acknowledges gratefully financial support of Polish
Ministry of Science and Higher Education (Grant No N N202 174735,
contract number 1747/B/H03/2008/35).

\end{acknowledgments}


\begin{thebibliography}{99}
\bibitem{BUB07A} A. Bubak, A. Budzanowski, D. Filges, F. Goldenbaum,
                 A. Heczko, H. Hodde, L. Jarczyk, B. Kamys, M.
                 Kistryn, St. Kistryn, St. Kliczewski,  A. Kowalczyk,
                 E. Kozik, P. Kulessa,  H. Machner,  A. Magiera,  W. Migda³,
                 N. Paul,  B. Piskor-Ignatowicz, M. Pucha³a,  K. Pysz,  Z. Rudy,
                 R. Siudak,  M. Wojciechowski, and P. W\"ustner,
                 Phys. Rev. C
                 \textbf{76},014618 (2007)
%
\bibitem{BUD08A} A. Budzanowski, M. Fidelus, D. Filges, F. Goldenbaum, H. Hodde, L. Jarczyk,
                 B. Kamys, M. Kistryn, St. Kistryn, St. Kliczewski, A. Kowalczyk,
                 E. Kozik, P. Kulessa, H. Machner, A. Magiera, B. Piskor-Ignatowicz,
                 K. Pysz,  Z. Rudy,  R. Siudak,  and M. Wojciechowski,
                 Phys. Rev. C \textbf{78}, 024603 (2008)
%
\bibitem{BAR04A} R. Barna, V. Bollini, A. Bubak, A. Budzanowski, D. D.Pasquale,
                 D. Filges, S. V. F\"ortsch, F. Goldenbaum, A. Heczko, H. Hodde,
                 A. Italiano, L. Jarczyk, B. Kamys, J. Kisiel, M.Kistryn,
                 St. Kistryn, St. Kliczewski, A. Kowalczyk, P. Kulessa,
                 H.Machner, A. Magiera, J.Majewski, W.Migda{\l}, H.Ohm, N.Paul,
                 B. Piskor-Ignatowicz, K. Pysz, Z. Rudy, H. Schaal, R. Siudak,
                 E. Stephan, G.F.Steyn, R.Sworst, T.Thovhogi, M.Wojciechowski,
                 W.Zipper,
                 Nucl. Instr. Meth. in Phys. Research A \textbf{519},
                 610 (2004)
%
\bibitem{BUB04A} A. Bubak, B. Kamys, M. Kistryn, B. Piskor-Ignatowicz,
                 Nucl. Instr. Meth. in Phys. Research B
                 \textbf{226}, 507 (2004)
\bibitem{HER06A} C.-M. Herbach, D. Hilscher, U. Jahnke, V.G. Tishchenko, J. Galin, A. Letourneau,
                 A. P\'eghaire, D. Filges, F. Goldenbaum, L. Pie\'nkowski, W.U. Schr\"oder, and J. T\"oke,
                 Nuclear Physics A \textbf{765}, 425 (2006)
%
\bibitem{RAI75A} G.M. Raisbeck, P. Boerstling, R. Klapisch, and T.D. Thomas,
                 Physical Review C \textbf{12}, 527(1975)
%
\bibitem{BOU04A} A. Boudard, J. Cugnon, S. Leray, and C. Volant,
                 Nucl. Phys. A \textbf{740}, 195 (2004)
%
\bibitem{FUR00A} S. Furihata, Nucl. Instr. and Meth. in Phys. Research B
                 \textbf{71}, 251 (2000)
\bibitem{FUR02A} S. Furihata and T. Nakamura, Journal of Nuclear Science
                 and Technology Supplement \textbf{2}, 758 (2002)
%
%
\bibitem{LET02A} A. Letourneau, A. B\"ohm, J. Galin, B. Lott, A. P\'eghaire,
                 M. Enke, C. M. Herbach, D. Hilscher, U. Jahnke, V. Tishchenko,
                 \emph{et al.}, Nucl. Phys. A \textbf{712}, 133 (2002)
%
\bibitem{WES76A} G.D. Westfall, J. Gosset, P.J. Johansen, A.M. Poskanzer,
                 W.G. Meyer, H.H. Gutbrot, A. Sandoval and R.Stock,
                 Phys. Rev. Lett. \textbf{37}, 1202 (1976)
%
%
\bibitem{WES78A} G. D. Westfall, R. G. Sextro, A. M. Poskanzer,
                 A. M. Zebelman, G. W. Butler, and E. K. Hyde,
                 Phys. Rev. C \textbf{17}, 1368 (1978)

%
%
%
%
%
%
%
%
%
%
%
%
%
%
%
%
%
\end{thebibliography}
\end{document}